\documentclass[aps,prl,twocolumn,amsmath,showpacs,amssymb,superscriptaddress]{revtex4}
\usepackage{graphicx}
\usepackage{dcolumn}
\usepackage{dcolumn}
\usepackage{color}
\allowdisplaybreaks

\begin{document}
\title{Standard Quantum Limit for Probing Mechanical Energy Quantization}
\author{Haixing Miao}
\affiliation{School of Physics, University of Western Australia, WA 6009, Australia}
\author{Stefan Danilishin}
\affiliation{Physics Faculty, Moscow State University, Moscow 119991, Russia}
\affiliation{Max-Planck Institut f\"ur Gravitationsphysik (Albert-Einstein-Institut)
and Leibniz Universit\"at Hannover, Callinstr. 38, 30167 Hannover, Germany}
\author{Thomas Corbitt}
\affiliation{LIGO Laboratory, NW22-295, Massachusetts Institute of Technology, Cambridge, MA 02139, USA}
\author{Yanbei Chen}
\affiliation{Theoretical Astrophysics 130-33, California Institute of Technology,
Pasadena, CA 91125, USA}
\affiliation{Max-Planck Institut f\"ur Gravitationsphysik (Albert-Einstein-Institut),
Am M\"uhlenberg 1, 14476 Golm, Germany}
\newcommand{\be}{\begin{equation}}
\newcommand{\ee}{\end{equation}}
\newcommand{\ba}{\begin{eqnarray}}
\newcommand{\ea}{\end{eqnarray}}
\newcommand{\R}[1]{\textcolor{red}{#1}}
\newcommand{\B}[1]{\textcolor{blue}{#1}}

\begin{abstract}
We derive a standard quantum limit for probing mechanical energy quantization in a class of systems
with mechanical modes parametrically coupled to external degrees of freedom. To
resolve a single mechanical quantum, it requires a strong-coupling regime --- the decay rate of external
degrees of freedom is smaller than the parametric coupling rate. In the case for cavity-assisted optomechanical systems, e.g. the one
proposed by Thompson {\it et al.} \cite{Thompson}, zero-point motion of the mechanical oscillator needs to be
comparable to linear dynamical range of the optical system which is characterized by the
optical wavelength divided by the cavity finesse.
\end{abstract}

\maketitle

{\it Introduction.}---Recently, significant cooling of mechanical modes of harmonic oscillators has been achieved by
extracting heat through parametric damping or active feedback~\cite{Thompson, Naik}. Theoretical
calculations suggest that oscillators with a large thermal occupation number
($k_B T \gg \hbar\omega_m$) can be cooled to be close
to their ground state, if they have high enough quality factors~\cite{Marquart}. Once the ground
state is approached, many interesting studies of macroscopic quantum mechanics can be performed,
e.g. teleporting a quantum state onto mechanical degrees of freedom
\cite{Mancini}, creating quantum entanglement between a cavity mode and an oscillator \cite{Vitali}
and between two macroscopic test masses~\cite{Helge}. Most proposals
involve the oscillator position linearly coupled to photons, in which case  the quantum features of the
oscillator, to a great extent, are attributable to the quantization of photons. In order to probe
the intrinsic quantum nature of an oscillator, one of the most transparent approaches is to directly
measure its energy quantization, and  quantum jumps between discreet energy eigenstates.
Since linear couplings alone will not project an oscillator onto its energy eigenstates, nonlinearities
are generally required \cite{Santamore, Martin, Jacobs}. For cavity-assisted optomechanical systems,
one experimental scheme, proposed in the pioneering work of Thompson {\it et al.} \cite{Thompson}, is
to place a dielectric membrane inside a high-finesse Fabry-Perot cavity, forming a pair of coupled
cavities~\footnote{A similar configuration has been proposed by Braginsky {\it et al.} for detecting gravitational-waves,
Phys. Lett. A {\bf 232}, 340 (1997) and Phys. Lett. A {\bf 246}, 485 (1998). }.
If the membrane is appropriately located, a dispersive coupling between the membrane position and
the optical field is predominantly quadratic, allowing the detection of mechanical energy quantization.

In this letter,  we show that in the experimental setup of Thompson {\it et al.}, the optical
field also couples linearly to the membrane. Due to finiteness
of cavity finesse (either intentional for readout or due to optical losses), this linear coupling
introduces quantum back-action. Interestingly, it sets forth a simple standard quantum limit,
which dictates that only those systems
whose cavity-mode decay rates are smaller than the optomechanical coupling rate can successfully 
resolve energy levels. We will  further show that a similar constraint applies universally to all 
experiments that attempt to probe mechanical energy quantization via parametric coupling with external 
degrees of freedom (either optical or electrical).

\begin{figure}
\includegraphics[width=8.6cm, bb=2 2 356 103,clip]{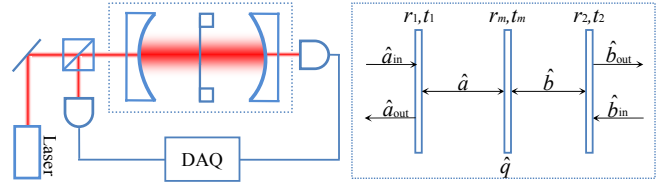}\caption{
The left panel presents the schematic configuration of coupled cavities in the proposed
experiment \cite{Thompson}. The right panel shows optical modes and we denote reflectivity and transimisivity of the optical elements by $r_i$ and $t_i\,(i=1,2,m)$.}
\label{config}
\end{figure}

{\it Coupled Cavities.}---Optical configuration of coupled cavities is shown in Fig.~\ref{config}. Given
the specification in Ref. \cite{Thompson}, transmissivities of the membrane and end mirrors are quite low, and
thus a two-mode description is appropriate \cite{Bhattacharya,Jayich}, with the corresponding Hamiltonian
\begin{align}
\label{1}\nonumber
\hat{\cal H}&=\hbar\,\omega_m(\hat{q}^2+\hat{p}^2)/2+\hbar\,\omega_0(\hat{a}^{\dag}
\hat{a}+\hat{b}^{\dag}\hat{b})-\hbar\,\omega_s(\hat{a}^{\dag}\hat{b}+
\hat{b}^{\dag}\hat{a})\\&+\hbar\, G_0\hat{q}(\hat{a}^{\dag}\hat{a}-\hat{b}^{\dag}\hat{b})
+\hat{\cal H}_{\rm ext}+\hat{\cal H}_{\xi}.
\end{align}
Here $\hat{q}, \hat{p}$ are normalized position and momentum of the membrane;
$\hat{a}, \hat{b}$ are annihilation operators of cavity modes in the individual cavities (both
resonate at $\omega_0$);
$\omega_s\equiv t_m c/L$ is the optical coupling constant for $\hat{a}$ and $\hat{b}$, through transmission
of the membrane~\cite{Jayich}; $G_0\equiv 2\sqrt{2}\omega_0x_q/L$ is the optomechanical coupling
constant with $L$ denoting the cavity length and zero-point motion $x_q\equiv\sqrt{\hbar/(2m\,\omega_m)}$;
$\hat{\cal H}_{\rm ext}$ and $\hat{\cal H}_{\xi}$ correspond to the coupling of the system to the
environment and quantify the fluctuation and dissipation mechanism.
By introducing {\it optical} normal modes, namely the common mode $\hat{c}
\equiv (\hat{a}+\hat{b})/\sqrt{2}$ and differential mode $\hat{d}\equiv(\hat{a}-\hat{b})/\sqrt{2}$,
\ba
{\hat{\cal H}}/\hbar &=&\frac{\omega_m}{2}(\hat{q}^2+\hat{p}^2)+\omega_{-}\hat{c}
^{\dag}\hat{c}+\omega_{+}\hat{d}^{\dag}\hat{d}+G_0\hat{q}(\hat{c}^{\dag}\hat{d}+\hat{d}^{\dag}\hat{c})
\nonumber\\&+& i(\sqrt{2\gamma_c}\,\hat{c}^{\dag}\hat{c}_{\rm in}+\sqrt{2\gamma_d}\,\hat{d}^{\dag}\hat{d}_{\rm in}-{\rm H.c.})+\hat{\cal H}_{\xi}/\hbar
\label{2}
\ea
where $\omega_{\pm}\equiv\omega_0\pm\omega_s$ and in the Markovian approximation 
$\hat{\cal H}_{\rm ext}$ is written out explicitly in
the second line (with $\gamma_{c,d}$ denoting decay rates and H.c. for Hermitian conjugate).

Before analyzing the detailed dynamics, here we follow Thompson {\it et al.}~\cite{Thompson} and Bhattacharya
and Meystre~\cite{Bhattacharya} by assuming $\omega_m \ll \omega_s$ and $G_0 \ll |\omega_+-\omega_-| =2\omega_s$, analogous
to the dispersive regime in the photon-number counting experiment with a superconducting qubit \cite{Schuster,Clerk}.
This allows us to treat $\hbar\, G_0\hat{q}(\hat{c}^{\dag}\hat{d}+\hat{d}^{\dag}\hat{c})$ as a perturbation and
diagonalize the Hamiltonian formally.
Up to $G_0^2/(2\omega_s)^2$, the optical and optomechanical coupling parts of the original
Hamiltonian can be written as
\be
\hat{\cal H}/\hbar=\left(\omega_{-}-\frac{G_0^2\hat{q}^2}{2\omega_s}\right)\hat{o}^{\dag}\hat{o}
+\left(\omega_{+}+\frac{G_0^2\hat{q}^2}{2\omega_s}\right)\hat{e}^{\dag}\hat{e}.
\label{3}
\ee
At first sight, frequency shift of the eigenmodes $\hat{o}$ and $\hat{e}$
is proportional to $\hat{q}^2$. Since frequency separation of two normal modes
is $2\,\omega_s\gg \gamma_{c,d}$, they can be independently driven and detected.
Besides, with $\gamma_{c,d}<\omega_m$, only averaged membrane motion
is registered and $\overline{\hat q^2}=\hat N+1/2$ with $\hat N$
denoting the number of quanta. Therefore, previous authors had concluded that such
a {\it purely} dispersive coupling allows quantum non-demolition (QND) measurements of
the mechanical quanta.

However, the new eigenmodes $\hat{o}$ and $\hat{e}$ are given by
\be
\hat{o}= \hat{c}-[({G_0 \hat{d}})/({2\omega_s})]\hat{q},\quad
\hat{e}= \hat{d}+[({G_0 \hat{c}})/({2\omega_s})]\hat{q}.
\label{4}
\ee
If we pump $\hat c$ with classical amplitude $\bar c$ and left $\hat d$ in vacuum state,
the detected mode $\hat o$ will have a negligible linear response. However, the idle mode
$\hat e\approx [G_0\bar c/(2\omega_s)]\hat q$, which is dominated by linear coupling. If we choose
to drive $\hat d$, the role of $\hat o$ and $\hat e$ will simply swap.
Such linear coupling can potentially demolish the energy eigenstates that we wish to probe.
We can make an order-of-magnitude estimate.
The optomechanical coupling term in Eq.~\eqref{2}, at the linear order,
reads $G_0 \hat q(\bar c\,\hat d + \bar c^* \hat d^\dagger)$.
According to the Fermi's golden rule, it causes decoherence of energy eigenstate near the ground level at a rate of
\be
\tau_{\rm dec}^{-1}=G_0^2|\bar c|^2\tilde S_{\hat d}(-\omega_m)\approx G_0^2|\bar c|^2\gamma_d/(2\omega_s^2),
\label{5}
\ee
where we have assumed that $\hat{c}$ is on resonance, and
\be
\tilde S_{\hat d}\equiv\mbox{$\int$} dt\,e^{i\omega t}\langle\hat d(t)\hat d^{\dag}(0) \rangle=
{2\gamma_d}/[{{(\omega-2\omega_s)^2+\gamma_d^2}}].
\label{6}
\ee
On the other hand, from Eq. \eqref{3} and linear response theory \cite{Clerk2}, the measurement
time scale to resolve the energy eigenstate (i.e. measuring $\hat N$ with a unit error) with a shot-noise
limited sensitivity is approximately given by
\be
\tau_{\rm m}\approx[{\gamma_c^2\omega_s^2}/({G_0^4|\bar{c}|^2})]\tilde S_{\hat c}(0)={2\omega_s^2\gamma_c}/({G_0^4|\bar{c}|^2}),
\label{7}
\ee
where $\tilde S_{\hat{c}}(0)$ is the spectral density of $\hat{c}$ at zero frequency. Requiring $\tau_{\rm m} \le\tau_{\rm dec}$ yields
\be
({\gamma_c\gamma_d}/{G_0^2})\lesssim1.
\label{7.1}
\ee
In the case when transmissivity of end mirrors $t_1=t_2\equiv t_0$, we have $\gamma_c=\gamma_d=c\,t_0^2/(2L)$.
Defining the cavity finesse as ${\cal F}\equiv\pi/t_0^2$, the above inequality reduces to
${\lambda}/({{\cal F} x_q})\lesssim 8\sqrt{2}$.
Therefore, to probe mechanical energy quantization, it requires a strong-coupling regime (c.f. Eq. \eqref{7.1}),
or equivalently, for such an optomechanical system, zero-point mechanical motion $x_q$ to be comparable
to linear dynamical range $\lambda/{\cal F}$ of the cavity.

We now carry out a detailed analysis of the dynamics according to the
standard input-output formalism \cite{Gardiner}. In the rotating frame at the laser frequency $\omega_+$,
the nonlinear quantum Langevin equations are given by
\begin{align}
\dot{\hat{q}}&=\omega_m\,\hat{p},\label{9}\\
\dot{\hat{p}}&=-\omega_m\,\hat{q}-\gamma_m\,\hat{p}-G_0(\hat{c}^{\dag}\hat{d}+\hat{d}
^{\dag}\hat{c})+\xi_{\rm th},\label{10}\\
\dot{\hat{c}}&=-\gamma_c\,\hat{c}-i\,G_0\,\hat{q}\,\hat{d}+\sqrt{2\gamma_c}\,\hat{c}_{\rm in},\label{11}\\
\dot{\hat{d}}&=-(\gamma_d+2\,i\,\omega_s)\,\hat{d}-i\,G_0\,\hat{q}\,\hat{c}+\sqrt{2\gamma_d}\,
\hat{d}_{\rm in}.\label{12}
\end{align}
Here the mechanical damping and associated Brownian thermal force $\xi_{\rm th}$ origin from $\hat {\cal H}_{\xi}$ under the
Markovian approximation.
These equations can be solved perturbatively by decomposing every Heisenberg operator $\hat{\alpha}$
into different orders such that $\hat{\alpha}=\bar{\alpha}+ \epsilon\, \hat{\alpha}^{(1)}+ \epsilon^2\hat{\alpha}^{(2)}+{\cal O}[\epsilon^3]$. We treat $G_0/(2\omega_s)$, vacuum fluctuations
$\sqrt{2\gamma_c}\,\hat{c}_{\rm in}^{(1)}$ and $\sqrt{2\gamma_d}\,\hat{d}_{\rm in}^{(1)}$ (simply denoted by $\sqrt{2\gamma_c}\,\hat{c}_{\rm in}$ and
$\sqrt{2\gamma_d}\,\hat{d}_{\rm in}$ in later discussions) as being of the order of $\epsilon$ ($\epsilon \ll 1$).

To the zeroth order, $\bar{c}=\sqrt{2I_0/(\gamma_c\hbar\,\omega_0)}$ with $I_0$ denoting
the input optical power and $\bar{d}=0$. Up to the first order, the radiation pressure term 
reads $G_0\bar{c}[\hat{d}^{(1)}+\hat{d}^{(1)\dag}]$ ($\bar c$ is set to be real
by choosing an appropriate phase reference). In the frequency domain, it can be written as
\ba
\label{14}
\tilde{F}_{\rm rp}=\frac{2\sqrt{\gamma_d}\,G_0\,\bar{c}[(\gamma_d-i\omega)\tilde{v}_1
-2\omega_s\tilde{v}_2]+4G_0^2\bar{c}^2\omega_s\tilde{q}}{(\omega+2\omega_s+i\gamma_d)(\omega-2\omega_s+i\gamma_d)},
\ea
where $\tilde{v}_1, \tilde{v}_2$ and $\tilde q$ are Fourier transformations of $\hat{v}_1(t)
\equiv (\hat{d}_{\rm in}+\hat{d}_{\rm in}^{\dag})/\sqrt{2}$, $\hat{v}_2(t)\equiv(\hat{d}_{\rm in}
-\hat{d}^{\dag}_{\rm in})/(i\sqrt{2})$ and $\hat{q}(t)$ respectively. The part, containing vacuum fluctuations,
is the back-action $\hat F_{\rm BA}$, which induces the quantum limit. The other part proportional to $\tilde{q}$ is the optical-spring effect.
Within the time scale for measuring energy quantization, of the order of $\gamma_c^{-1}\,(\ll\gamma_m^{-1})$,
the positive damping can be neglected but the negative rigidity has an interesting consequence ---
it modifies $\omega_m$ to an effective $\omega_{\rm eff}\,(<\omega_m)$. Correspondingly,
position of the high-Q membrane is
\be
\hat{q}(t)=\hat{q}_m+\Lambda^2\mbox{$\int_0^t$}dt'\sin\omega_{\rm eff}(t-t')[\hat F_{\rm BA}(t')+\xi_{\rm th}(t')]
\label{16}
\ee
with $\Lambda\equiv\sqrt{\omega_m/\omega_{\rm eff}}$. The free quantum oscillation $\hat{q}_m=\Lambda\,(\hat{q}_0\cos\omega_{\rm eff}\,t+\hat p_0\sin \omega_{\rm eff}\,t)$
and $\hat{q}_0$ and $\hat{p}_0$ are the initial position and momentum normalized with respect to $\sqrt{\hbar/(m\,\omega_{\rm eff})}$ and $\sqrt{\hbar\,m\,\omega_{\rm eff}}$.

The dispersive response is given by the second-order perturbation ${\cal O}[\epsilon^2]$.
Adiabatically eliminating rapidly oscillating components and assuming $\omega_m\ll \omega_{s}$ which
can be shown to maximize the signal-to-noise ratio, we obtain
\begin{align}\nonumber
\hat{c}^{(2)}(t)&=-iG_0\mbox{$\int_0^t$} dt' e^{-\gamma_c(t-t')}\hat{q}(t')\,\hat{d}^{(1)}(t')\\
&\approx G_{\rm eff}^2\,\bar{c}\,\hat{N}(t)/(2i\gamma_c\,\omega_s)\,.
\label{15}
\end{align}
Here $G_{\rm eff}\equiv \Lambda\,G_0$ and $\hat N(t)\equiv
\hat N_0+\Delta\hat N(t)$ contains the number of mechanical quanta $\hat N_0\equiv (\hat q_0^2+\hat p_0^2)/2$ and
the noise term $\Delta\hat N(t)$ due to the back-action and thermal noise.
To read out $\hat N(t)$, we integrate output phase quadrature for a duration $\tau$. According to the input-output relation
$\hat{c}_{\rm out}+\hat{c}_{\rm in}=\sqrt{2\gamma_c}\,\hat{c}$, the estimator reads
\be
\hat Y(\tau)=\mbox{$\int_0^{\tau}$}dt[\hat{u}_2(t)-{G_{\rm eff}^2\,\bar{c}}\,\hat N(t)/({\sqrt{\gamma_c}\,\omega_s})
],
\label{17}
\ee
where $\hat{u}_2\equiv (\hat{c}_{\rm in}-\hat{c}_{\rm in}^{\dag})/(i\sqrt{2})$. For
Gaussian and Markovian process, the correlation function $\langle\hat{c}_2(t)\,\hat{c}_2^{\dag}(t')\rangle=\delta(t-t')/2$. For
typical experiments, the thermal
occupation number $\bar{n}_{\rm th}\equiv k_B T/(\hbar\,\omega_m)$ is much larger than unity, and
$\langle\xi_{\rm th}(t)\,\xi_{\rm th}(t')\rangle\approx2\gamma_m \bar{n}_{\rm th}\,\delta(t-t')$.
Through evaluating the four-point correlation function of back-action noise
and $\xi_{\rm th}(t)$ in $\langle\Delta \hat N(t)\Delta \hat N(t')\rangle$,
we obtain the resolution $\Delta N$ as a function of $\tau$
\be
\Delta N^2=\left(\frac{\gamma_c\omega_s^2}{G_{\rm eff}^4\bar{c}^2\tau}\right)+\frac{5}{6}\left(\frac
{\gamma_d G_{\rm eff}^2\bar{c}^2\tau}{2\sqrt{2}\,\omega_s^2}\right)^2+\frac{5}{6}\left(\frac{\gamma_mk_B
T\tau}{\sqrt{2}\,\hbar\,\omega_{\rm eff}}\right)^2.
\label{18}
\ee

In order to successfully observe energy quantization, the following conditions are simultaneously required:
(i) the resolution $\Delta N^2$ should have a minimum equal or less than unity. (ii) this minimum should be
reachable within $\tau$ that is longer than the cavity storage time $1/\gamma_c$ (which in turn must be longer
than the oscillation period $1/\omega_{\rm eff}$ of the membrane). (iii) the system dynamics should be stable
when taking into account optical rigidity which is approximately equal to $G_0^2\bar{c}^2/\omega_s$ for $\omega_m\ll\omega_s$.

Specifically, the standard quantum limit in condition (i), set by the first two terms in $\Delta N^2$,
gives ${\gamma_c\gamma_d}/{G_{\rm eff}^2}\lesssim 1$, or equivalently
$({\gamma_c\gamma_d}/{G_0^2})\lesssim \Lambda^2.$
If we neglect the optical spring effect ($\Lambda =1$), we simply recover Eq. \eqref{7.1}.
A strong negative optical rigidity $(\omega_{\rm eff}\ll \omega_m, \,\mbox{i.e.}\,\Lambda\gg 1)$
can significantly enhance the effective coupling strength and ease the requirements on optomechanical properties.
However, a small $\omega_{\rm eff}$ also makes the system susceptible to the thermal noise. Taking account of all 
the above conditions, the optimal
$\omega_{\rm eff}=\omega_m\sqrt{\bar n_{\rm th}/Q_m}$ with mechanical quality factor $Q_m\equiv \omega_m/\gamma_m$,
and there is a nontrivial constraint on the thermal occupation number, which reads $
(\bar{n}_{\rm th}/Q_m)\le [G_0^2/(\omega_s\gamma_c)]^{2/3}.$
\begin{figure}
\includegraphics[width=0.45\textwidth, bb=0 0 500 231,clip]{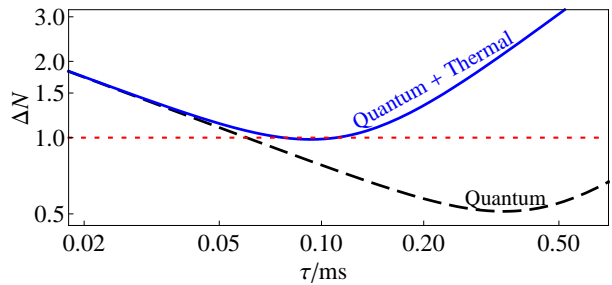}
\caption{The resolution $\Delta N$ for measuring mechanical energy quantization depending on the integration duration $\tau$
with total noise (Solid) and quantum noise only (Dashed).} \label{DN}
\end{figure}

For numerical estimate, we use a similar specification as given in Ref. \cite{Thompson} but assume a slightly higher mechanical
quality factor $Q_m$, lower environmental temperature $T$ and lower input optical power $I_0$ such that all mentioned
conditions are satisfied. The parameters are the following: $\,m=50\,{\rm pg},\,\omega_m/(2\pi)=10^5\,{\rm Hz}, \,Q_m=3.2\times 10^7,\lambda=532\,{\rm nm},\, L= 3\,{\rm cm},\,r_m=0.9999,\,{\cal F}=6\times 10^5,\, T=0.1\,{\rm K}$ and $I_0\approx 5$ nW.
The resulting resolution $\Delta N$ is shown in Fig. \ref{DN}, and
we are able to resolve single mechanical quantum when $\tau\approx 0.1$ ms.

Even thought we have been focusing on the double-sided setup where $t_1\approx t_2$,
the quantum limit also exists in the single-sided case originally proposed in Ref. \cite{Thompson}. Ideally,
a single-sided setup consists of a totally reflected end mirror and the vacuum fluctuations only enter from the
front mirror. Therefore, the quantum noises inside two sub-cavities have the same origin but different optical path.
Through similar input-output calculations, we find that if laser detuning is equal to $\pm\,\omega_s$,
the quantum noises destructively interfere with each other at low frequencies, due to the same mechanism
studied in great details in Ref. \cite{Elste}, achieving an ideal QND measurement. However, in
reality, the end mirror always has some finite transmission or optical loss which introduces uncorrelated vacuum
fluctuations. As it turns out, the quantum limit is similar to Eq. \eqref{7.1}, only with $\gamma_{c,d}$
replaced by the damping rate of two sub-cavities.

{\it General Systems.}---Actually,
the standard quantum limit obtained above applies to all schemes that attempt to probe mechanical energy quantization
via parametric coupling. Let us consider $n$ mechanical modes parametrically
coupled with $n'$ normal external modes, describable  by the following Hamiltonian
\ba
\hat{\cal H}&=&\sum_{\nu=1}^{n}\hbar\,\Omega
_{\nu}(\hat{q}^2_{\nu}+\hat{p}^2_{\nu})/2+\sum_{i=1}^{n'}\hbar\,\omega_{i}\,\hat a_i^{\dag}\hat a_i
\\\nonumber&+&\sum_{i,j=1}^{n'}\sum_{\nu=1}^{n}\hbar\,\chi_{ij
\nu}\,\hat{q}_{\nu}(\hat a_i^{\dag}\hat a_j+\hat a_j^{\dag}\hat a_i)+\hat{\cal H}_{\rm ext}+\hat{\cal H}_{\xi}\,.
\label{20}
\ea
Here Greek indices identify mechanical modes and Latin indices identify external modes; $\Omega_{\nu}$ and $\omega_i$
are eigenfrequencies; $\hat q_{\nu}, \hat p_{\nu}$ are normalized positions and momenta;
$\hat a_i$ are annihilation
operators of the external degrees of freedom; $\chi_{ij\nu}=\chi_{ji\nu}$ are coupling constants.
Similarly, we  focus on the regime where $|\chi_{ij\nu}|\ll|\omega_{i}-\omega_{j}|$
(dispersive) and $\Omega_{\nu}\ll|\omega_{i}-\omega_{j}|$ (adiabatic), and obtain
\be
\hat{\cal H}=\sum_{\nu=1}^{n}\hbar\,\Omega_{\nu}(\hat{q}^2_{\nu}+\hat{p}^2_{\nu})/2+\sum_{i=1}^{n'}\hbar\,\omega_{i}'
\hat{o}_i^{\dag}\hat{o}_i+\hat{\cal H}_{\rm ext}+\hat{\cal H}_{\xi},
\label{21}
\ee
where, up to $\chi_{ij{\nu}}^2/|\omega_i-\omega_j|^2$,
\be
\omega_i'=\omega_i+\sum_{\nu}\chi_{ii\nu}\hat{q}_{\nu}+\sum_{j\neq i}\sum_{\nu}\frac{(\chi_{ij{\nu}}
\hat{q}_{\nu})^2}{\omega_i-\omega_j}.
\label{22}
\ee
In order to have quadratic couplings between a pair of external and mechanical modes, $\hat{o}_{1}$
and $\hat{q}_1$ for instance, we require that $\chi_{11\nu}=0$ and $\chi_{1 i \nu}=\chi_{1 i 1}\delta_{1\nu}$, and then
\be
\omega_{1}'=\omega_{1}+\sum_{i\neq1} \frac{\chi_{1i1}^2}{\omega_{1}-\omega_{i}}\,\hat{q}_1^2.
\label{23}
\ee
However, there still are linear couplings which originate from idle modes. This is because, up to
$\chi_{ij{\nu}}/|\omega_i-\omega_j|$,
\be
\hat{o}_i=\hat{a}_i+\sum_{j\neq i}\frac{\chi_{ij1}\hat{a}_j}{\omega_i-\omega_j}\,
\hat{q}_1\approx\hat{a}_i+ \frac{\chi_{1i1}\bar{a}_{1}}{\omega_i-\omega_{1}}\,
\hat{q}_1 ~~~(i\neq 1).
\label{24}
\ee
where $\hat{a}_1$ is replaced with its classical amplitude $\bar{a}_1$, for $\bar{a}_1\gg \hat{a}_i$.
From Eq. \eqref{23} and \eqref{24}, both linear and dispersive couplings are inversely proportional
to $|\omega_i-\omega_1|$. Therefore, we only need to consider a tripartite system formed by $\hat q_1$, $\hat{o}_1$ 
and $\hat{o}_2$ which is the closest to $\hat{o}_1$ in frequency. The resulting Hamiltonian
is identical to Eq. \eqref{2}, and thus the same standard quantum limit applies.

{\it Conclusion.}--- We have demonstrated the existence of quantum limit for probing mechanical
energy quantization in general systems where mechanical modes parametrically interact with optical
or electrical degrees of freedom. 
This work will shed light on choosing the appropriate parameters for experimental realizations.

{\it Acknowledgements.}---We thank F. Ya. Khalili, H. M\"{u}ller-Ebhardt, H. Rehbein, and
our colleagues at TAPIR and MQM group for fruitful discussions. We also thank J. G. E. Harris and F. Marquardt for critical
comments and invaluable suggestions on the early manuscript. H. M. thanks D. G. Blair, L. Ju and C. Zhao for their keen
supports of his visits to Caltech where this work has been done. H. M. is supported by
the Australian Research Council and the Department of Education, Science and Training. S. D. and Y. C.
are supported by the Alexander von Humboldt Foundation's Sofja Kovalevskaja Programme, NSF grants
PHY-0653653 and PHY-0601459, as well as the David and Barbara Groce startup fund at Caltech. T. C.
is supported by NSF grants PHY-0107417 and PHY-0457264, and by the Sloan Foundation.

\end{document}